\documentclass[sigconf]{acmart}

\usepackage{multirow,color, hyperref, url, graphics}
\usepackage{listings}
\usepackage{xcolor}
\usepackage{amsmath}
\usepackage{diagbox}
\usepackage{geometry}
\usepackage{graphicx}
\usepackage{subfig}
\usepackage{balance}
\usepackage{adjustbox}
\usepackage{colortbl}
\usepackage{framed}

\newcommand{\tab}{\hspace*{1em}}

\setcopyright{none}

\renewcommand\footnotetextcopyrightpermission[1]{}

\acmDOI{}

\acmISBN{}

\acmConference[]{}{}{}
\acmYear{2021}
\copyrightyear{2021}

\acmArticle{}
\acmPrice{}

\setcopyright{acmcopyright}
\copyrightyear{2020}
\acmYear{2020}

\newcommand{\code}[1]{{\fontfamily{cmtt}\fontseries{m}\fontshape{n}\selectfont\small{#1}}}

\begin{document}

\lstset{
basicstyle=\ttfamily,
columns=fullflexible,
showstringspaces=false,
keywordstyle= \color{ blue!70},commentstyle=\color{red!50!green!50!blue!50},
frame=shadowbox,
rulesepcolor= \color{ red!20!green!20!blue!20}
}

\title{VM Matters: A Comparison of WASM VMs and EVMs \\
in the Performance of Blockchain Smart Contracts}

\author{Shuyu Zheng$^{1}$, Haoyu Wang$^{2}$, Lei Wu$^{3}$, Gang Huang$^{1}$, Xuanzhe Liu$^{1}$}
\affiliation{
  \institution{
  {\small{$^1$}}Peking University; {\small{$^2$}}Beijing University of Posts and Telecommunications; {\small{$^3$}}Zhejiang University
  }
}
\email{zhengshuyu@pku.edu.cn, haoyuwang@bupt.edu.cn, lei_wu@zju.edu.cn, {hg, liuxuanzhe}@pku.edu.cn}


\begin{abstract}
WebAssemly is an emerging runtime for Web applications and has been supported in almost all browsers. Recently, WebAssembly is further regarded to be a the next-generation environment for blockchain applications, and has been adopted by Ethereum, namely eWASM, to replace the state-of-the-art EVM. However, whether and how well current eWASM outperforms EVM on blockchain clients is still unknown. This paper conducts the first measurement study, to measure the performance on WASM VM and EVM for executing smart contracts on blockchain. To our surprise, the current WASM VM does not perform in expected performance. The overhead introduced by WASM is really non-trivial. Our results highlight the challenges when deploying WASM in practice, and provide insightful implications for improvement space. 
\end{abstract}


\maketitle

\section{Introduction}\label{sec:introduction}

Blockchain, as a trustable distributed ledger originated from Bitcoin~\cite{nakamoto2019bitcoin}, has been proliferating rapidly over the last few years. Bitcoin system has demonstrated that it is feasible to use the Internet to construct a decentralized value-transfer system that can be shared across the world and virtually free to use. Due to performance and scalability issues, however, it is difficult for Bitcoin to support large applications. Thus, Ethereum~\cite{ethereum} was proposed to allow users to develop smart contracts and create DApps that support advanced functionalities. Smart contract is designed as a computer protocol that allows users to digitally negotiate an agreement in a convenient and secure way. Compared with the traditional contract law, the transaction costs of a smart contract are dramatically reduced, and the correctness of its execution is ensured by the underlying consensus protocol.

The expansion of blockchain networks put forward higher requirements for scalability, security and efficiency. Dinh et al.~\cite{dinh2017blockbench} found that consensus protocols, transaction signing and execution engine are the main bottlenecks of Ethereum. The execution engine in the current Ethereum platform is called Ethereum Virtual Machine (EVM). Ethereum smart contracts are currently written in Solidity/Vyper and compiled down into EVM bytecode which is executed by every Ethereum node. As the smart contracts become more functional, the need of speeding up EVM execution has become more urgent. Further more, the tool and language support for writing smart contracts and compiling to EVM bytecode are not well-equipped for real-world applications~\cite{eth2.0, eth2.0explained}.

The developers of blockchain platforms turned their eyes to WebAssembly (WASM)~\cite{WebAssembly}. 
WASM is a new, low-level, binary format and runtime for the web, and has now been implemented in all major browsers~\cite{haas2017bringing,WebAssembly}. The WASM bytecode is designed to serve as a compilation target and has been supported by a wide range of modern programming languages since it was released officially in 2017~\cite{WasmLanguageList}. 
Previous studies~\cite{haas2017bringing,herrera2018webassembly} suggest that WASM programs exhibit near native speed, about 2x faster than their Javascript counterpart in web browsers. Apart from browsers, WASM is designed to support embeddings in memory-safe, sandboxed environment, for example, blockchains.

The Ethereum community has placed Ethereum flavored WebAssembly (eWASM) on the Ethereum 2.0 roadmap as a replacement to the first generation of EVM. The idea seems promising. Firstly, WASM is supported by a rich variety of programming languages, blockchain developers can benefit from the WASM ecosystem, rather than relying on Solidity/Vyper only. 
Secondly, more transactions can be put into a block if eWASM executes faster~\cite{eth2.0},
which may improve the scalability of Ethereum network. 
Till now, major Ethereum clients such as Go-ethereum~\cite{goethereum} and Openethereum~\cite{openethereum} have implemented experimental WASM supports.
Ethereum is not the first to use WASM virtual machine (WASM VM) as its contract execution engine. Many popular blockchain platforms, such as EOSIO~\cite{eosio} and NEAR~\cite{near}, are using execution engines based on WASM. Some others, like Tron~\cite{tronwasm} and Perlin~\cite{life}, are also on the way to develop their own WASM VMs.

While WASM is generally expected to improve the performance of blockchain execution engines~\cite{ewasmdesign, eth2.0, eth2.0explained}, there has been no clear evidence to support this statement. 
\textit{Whether WASM bytecode really outperforms EVM bytecode on blockchain clients is still unknown.} This is non-trivial in practice as the performance of VM can definitely impact the scalalibity of the whole blockchain applications. To the best of our knowledge, neither the research community nor the industry has investigated the potential performance improvement introduced by WASM VMs in a detailed manner.

\textbf{This Work.}
In this paper, we take the first step to measure the impact of contract engines on the performance of smart contract execution. To be specific, we focus on two types of VM engines, \textit{EVM engines} and \textit{WASM engines}, corresponding to the execution of EVM bytecode and WASM bytecode, respectively.
Our study covers almost all the popular engines EVM and WASM engines (\textbf{see Section~\ref{sec:studydesign}}), including two most widely used client-based EVM engines (\texttt{Go-Ethereum} and \texttt{Openethereum}), and 11 state-of-the-art WASM engines (including three engines specialized for blockchain, i.e., EOS VM engine, and WASM-version engines provided by \texttt{Go-Ethereum} and \texttt{Openethereum}).
To pinpoint the root cause leading to the performance bottleneck, we provide a comprehensive study that includes both \textit{intra-bytecode} engine comparison study and \textit{inter-bytecode} engine comparison study (\textbf{see Section~\ref{sec:results}}). 
For the intra-bytecode engine study, we compare the contract execution performance within EVM engines and within WASM engines respectively, which can provide insights on the performance overhead (bottleneck) caused in the design of some specific engines. 
For the inter-bytecode engine study, we compare the contract execution performance across the VM engines that support different types of bytecode. Specifically, we compare the EVM-version and WASM-version engines of \texttt{Go-Ethereum} and \texttt{Openethereum}, as they are by far the most popular blockchain clients that support these two types of engines. 
Considering that no benchmarks are available in the research community, we further make effort to design a set of benchmarks with both WASM bytecode and EVM bytecode formats for evaluation, covering the most representative operations in smart contracts. Among many interesting results and observations, the following are prominent: 

\begin{itemize}
\item \textit{The support for WASM on Ethereum-based blockchain clients is far from satisfactory.} We observe a number of emerging issues: 1) existing WASM support is unstable and much effort is needed to activate WASM engines; 2) it is impossible to implement eWASM smart contracts in a general form as the coding convention varies across clients; and 3) few languages are supported by existing tool chains, and the contract development process is cumbersome and error-prone.

\item \textit{For the two most popular Ethereum EVM engines, the performance of contract execution on Geth is better than that on Openethereum.} The opcode-level implementation of EVM engines would greatly impact the overall performance.

\item \textit{The performance of smart contract execution across standalone WASM engines differs greatly, even up to three orders of magnitude.} While WASM bytecode is generally expected to achieve excellent performance, the implementation of WASM engines can greatly impact the performance. 

\item \textit{The native supported data type has a great impact on the performance of WASM contract execution.} As the native data type of WASM is 32/64 bit, it would introduce additional overhead during the execution of 256-bit smart contracts (the native data type of EVM bytecode).

\item \textit{While standalone WASM engines are faster than EVM engines, the eWASM VMs are slower than EVMs for all of the 256-bit benchmarks and most of the 64-bit benchmarks on Geth and Openethereum clients.} This is because the overhead introduced by gas metering before the execution of WASM bytecode and the inefficient context switch of EEI methods.

\end{itemize}

To the best of our knowledge, this is the first measurement study in our research community that investigates the performance of smart contract execution across a number of EVM and WASM engines.
Our major observations, however, contradict the general expectation that WASM engines will improve the contract execution performance. 
The overhead introduced by gas metering and WASM VM embedding overwhelms the performance gain of WASM, thus, the WASM engines run slower than EVM engines on most of the benchmarks in blockchain clients Geth and Openethereum. There is a need for optimizing the execution process of WASM on the blockchain client and improving the interface design between WASM and the blockchain environment to reduce additional overhead. Our efforts highlight the practical challenges of applying WASM engines in executing smart contracts and promote better operational practices. To boost future research, we will release the crafted benchmark and all the experiment results to the research community (link removed due to anonymous submission).

\section{Background}\label{sec:background}

\begin{table*}[t]
\centering
\small
\caption{A Comparison of EVM, WASM and eWASM bytecode.}
\label{table:ewasm_design}
\begin{tabular}{llccc}
\hline
 & & \textbf{EVM} & \textbf{WASM} & \textbf{eWASM}\\ 
\hline
\hline
\textbf{Opcodes} & & \begin{tabular}[c]{@{}c@{}}Many high-level instructions \\ (\texttt{SSTORE}, \texttt{CREATE}, \texttt{CALL}, \texttt{SHA3} etc.)\end{tabular} & Low-level instructions & \begin{tabular}[c]{@{}c@{}}A subset of WASM. High-level instructions \\ are defined through Ethereum Environment \\ Interface (EEI) implemented by client engines. \end{tabular}\\ 
\hline
\multirow{2}{*}{\textbf{Data types}} & \textbf{Integers} & 8 up to 256 bits & 32/64 bit & 32/64 bit \\ 
\cline{2-5} 
& \textbf{Floating Point} & Not fully supported & 32/64 bit & Not supported \\ 
\hline
\textbf{Nondeterminism} & & N & Y & N \\
\hline
\end{tabular}
\end{table*}

We start by briefly presenting key concepts, including smart contract, and EVM/WASM/eWASM bytecode.

\subsection{Smart Contract}
Smart contract is designed as a computer protocol that allows users to digitally negotiate an agreement in a convenient and secure way. In the context of blockchain, the smart contract stands for complex applications having digital assets being directly controlled by a piece of code~\cite{ethereum}. The smart contracts are like autonomous agents deployed on the chain identified by a unique address, containing money balance in ether. Every time the contract account receives a message (from a contract) or transaction (from an external actor), it executes a specific piece of code, controlling over its own ether balance and key/value store to keep track of persistent variables. To interact with a contract, a user will submit a transaction carrying any amount of gas and a data payload formatted according to the Contract Application Binary Interface (ABI).

Ethereum acts as the first blockchain system that supports developing smart contract. In Ethereum, smart contracts are typically written in higher level languages, e.g., Solidity and Vyper (languages similar to JaveScript and Python), and then compiled to Ethereum Virtual Machine (EVM) bytecode.
The EVM bytecode will be executed in a virtual machine (VM) embedded inside of the Ethereum execution environment. 
The code execution is an infinite loop that repeatedly carries out the operation at the current program counter starting at the beginning of the bytecode, and then increases the program counter by one. The execution finales when the end of the code is reached or an error occurs or \texttt{STOP} or \texttt{RETURN} instruction is detected.
Users have to pay for every resource that they consume, including computation, bandwidth and storage. The computation cost of the smart contract is represented in the fundamental unit ``gas''. The calculation of gas takes place before each operation. If the VM finds that there is not enough gas, execution is halted and the transaction is reverted~\cite{ethereum}.

Note that, smart contracts in different blockchains do not necessarily follow the EVM bytecode format. As aforementioned, EOSIO, as the first blockchain that accepts Delegated Proof-of-Stake (DPoS) consensus protocol, adopts C++ as the official language to develop smart contracts. In particular, the source code of smart contract is first compiled down to WASM bytecode. Upon invocation, the bytecode will then be executed in EOSIO's WASM VM, resulting in transactions recorded on the blockchain. A number of popular blockchains including Ethereum and Tron, have claimed to support WASM bytecode in their next generation VM engines.

\subsection{EVM/WASM/eWASM Bytecode}
We next introduce the similarities and differences in the design and implementation of EVM, WASM and eWASM bytecode.  Table~\ref{table:ewasm_design} provides a high-level summary.

\subsubsection{EVM Bytecode}
EVM stands for Ethereum Virtual Machine. The Ethereum smart contract is compiled to a low-level, stack-based bytecode language called \textit{EVM bytecode}. 
The code consists of a series of operations. There are three types of space to store data, including stack, memory and long-term storage. The stack and the memory are reset after computation ends, while the storage persists for the long term.
The EVM has many high-level operations, such as \texttt{SSTORE} (writes a 256-bit integer to storage), \texttt{CREATE} (creates a child contract) and \texttt{CALL}(calls a method in another contract), \texttt{SHA3} (compute Keccak-256 hash). The native data type of EVM is 256-bit integer. It supports down to 8-bit integer calculation. But computations smaller than 256-bit must be converted to a 256-bit format before the EVM can process them. To ensure the determinism, EVM does not fully support floating point. Floating point can be declared in the smart contract, but cannot be assigned to or be used in the calculation.
Although there are ways to optimize EVM execution via just-in-time compilation, the existing implementations of EVM are rather primitive.

\subsubsection{WASM}
The Web has become the most ubiquitous platform for running all kinds of applications. JavaScript, the only natively supported programming language on the Web, cannot meet the growing demand of CPU intensive Web applications, such as interactive 3D visualization and games. In 2017, a group of browser vendors addressed this situation~\cite{haas2017bringing} by putting forward a safe, fast, portable low-level compilation target called WebAssembly. WebAssembly, aka \emph{WASM}, 
is a binary instruction format for a stack-based virtual machine that features near-native execution performance. To ensure security, WASM executes within a sandboxed environment separated from the host runtime, and enforces the security policies for permissions~\cite{WebAssembly}.
WASM is now supported by major browsers~\cite{WebAssembly}.
A growing number of programming languages~\cite{WasmLanguageList,reiser2017accelerate} including C/C++/C\#/.Net/Java/Go/Lua, can be compiled to WASM bytecode.

The semantics of WASM is divided into three phases: decoding, validation and execution. The execution phase is done within the embedding environment on a virtual machine. The WASM abstract machine includes a stack recording operand values and control constructs, and an abstract store containing the global state. It manipulates values of the 4 basic value types: integers and floating-point data with both 32-bit and 64-bit support, respectively.

\subsubsection{eWASM} \label{sec:background_ewasm}
eWASM~\cite{ewasmdesign} stands for Ethereum flavored WebAssembly. 
It is a restricted subset of WASM to be used for contracts in Ethereum. Since WASM does not support high-level opcodes, interaction with the Ethereum environment are defined through Ethereum Environment Interface (EEI) implemented by WASM execution engines. eWASM restricts non-deterministic behavior such as floating point, because every node in the blochain has to run with complete accuracy. The metering of computation cost in eWASM is similar to that of EVM. The WASM execution engine sums up the costs on the execution of each operation. The eWASM project also defines an EVM transcompiler to achieve backwards compatibility with the EVM bytecode on the current blockchain.

\section{Study Design}
\label{sec:studydesign}
In this section, we present the details of our measurement study of EVM and WASM engines. We first describe our research questions, then present the overall study design, followed by the brief introduction of our selected engines, crafted benchmark and the experimental setup.

\begin{figure*}[t!]
	\includegraphics[width=0.99\textwidth]{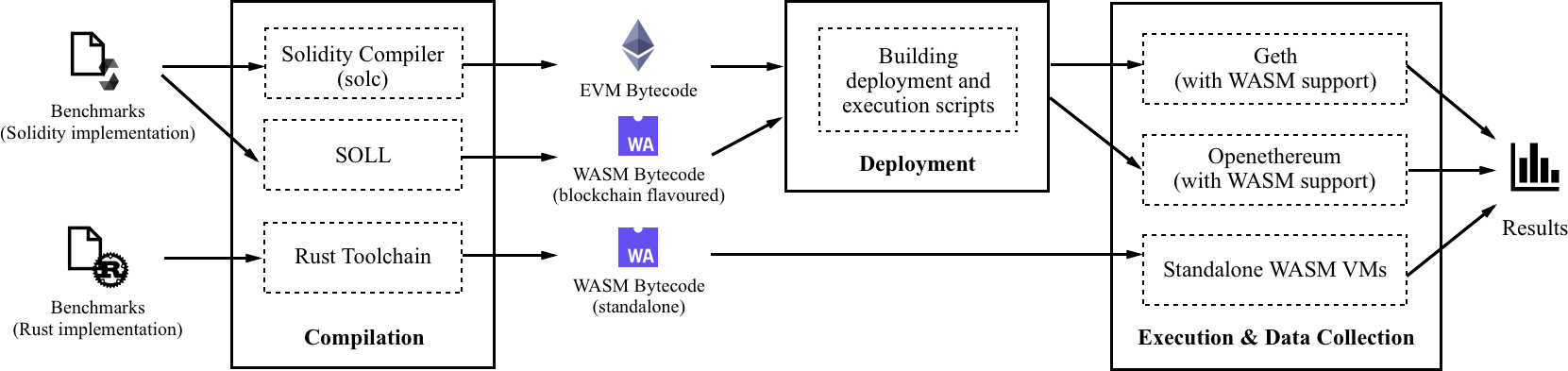}
	\caption{An overview of our measurement study.}
	\label{figure:overview}
\end{figure*}

\subsection{Research Questions}
We seek to focus on the following three research questions (RQs):

\begin{itemize}
    \item[RQ1] \textbf{WASM supports of Blockchain Clients.} Considering that major Ethereum clients have claimed to support WASM bytecode in order to embrace EVM 2.0, it is thus interesting to investigate \textit{how well do existing Ethereum-based blockchain clients support WASM bytecode?}
    \item[RQ2] \textbf{A Comparison of EVM Engines.} As there are several clients that support the execution of EVM bytecode, we are wondering \textit{are there any performance gaps of running smart contracts among them?} The result can provide insights on analyzing the performance overhead caused by the design of some specific engines. 
    \item[RQ3] \textbf{A Comparison of WASM Engines.}
    Similar to RQ2, there are a number of WASM engines available, including both client-based blockchain engines and standalone engines. We seek to explore \textit{to what extent does the performance gap exist among the WASM engines?}
    \item[RQ4] \textbf{WASM Engines VS. EVM Engines}
    It is widely acknowledged that WASM bytecode has the advantage of execution performance. Thus, it is interesting to investigate \textit{whether the execution performance of WASM bytecode outperforms EVM bytecode on existing blockchain clients? If so, how large is the performance gap can we expect?}
\end{itemize}

\subsection{Overview}

Figure~\ref{figure:overview} shows an overview of our measurement study. First, we need to create a comprehensive benchmark that provide both EVM bytecode and its equivalent WASM bytecode to enforce fair comparison. 
Note that, for the smart contract written in Solidity, we take advantage of Solidity Compiler solc (v0.5.4) to compile it to EVM bytecode. We further utilize SOLL~\cite{soll}, a newly emerging compiler for generating eWASM bytecode from Solidity. 
However, the eWASM bytecode cannot work on other general WASM engines. Thus, for a given solidity smart contract, we have to implement an equivalent contract using WASM-supported high level languages, and then compile it to WASM bytecode. In this paper, we make effort to implement the corresponding benchmark samples in Rust, and take advantage of Rust toolchain (nightly-2018-11-12) to get the WASM bytecode. 

Next, we build the deployment and execution scripts with the compiled EVM bytecode or eWASM bytecode. The deployment scripts are executed once to deploy the smart contracts on the blockchain. The execution script call the deployed smart contract without creating a transaction on the block chain. The scripts are run on a console connected to Ethereum private chain through IPC pipe. Note that, the eWASM bytecode compiled by SOLL cannot be deployed and run directly on Openethereum client, since the coding convention of WASM is different between Geth and Openethereum (will be further explained in Section~\ref{sec:results_support}). We thus manually transform the entry point, export functions and library functions of the compiled eWASM bytecode to meet the Openethereum WASM coding convention.

After that, we measure the execution time of EVM and WASM by instrumenting the blockchain clients. Specifically, we insert the time measuring code snippet before and after the contract execution loop on Geth and Openethereum, respectively. Besides, we also instrument the standalone WASM VMs before the decoding phase and after the execution phase.
Finally, we conduct the execution experiments on Geth, Openethereum and standalone WASM VMs and collect contract execution time. On Geth and Openethereum, we invoke a message \textit{call} transaction, which is directly executed in the VM of the node, but never mined into the blockchain.

\subsection{Selected Engines}
Our study seeks to cover all the most popular EVM and WASM engines. To be specific, we select 2 Ethereum clients with WASM support and 10 most popular WASM non-web embeddings on Github to perform comparison.

\subsubsection{Selected EVMs}
Two most popular Ethereum clients with WASM support are selected as the targets.
\begin{itemize}
    \item[1)] \textbf{Go-Ethereum}~\cite{goethereum} is the official Golang implementation of the Ethereum protocol. It is one of the most popular Ethereum implementations available either as a standalone client called \textit{\textbf{Geth}} that installs on operating systems, or as a library. We use the Geth client to deploy and execute the compiled smart contracts. We insert code before and after the execution of EVM to get the execution time. The experiments are performed on Geth version 1.9.1 with Go version 1.14.1.
    \item[2)] \textbf{Openethereum}~\cite{openethereum} was known as the "Parity Ethereum", another version of client of Ethereum. The ownership of Parity codebase and maintenance was transitioned and Parity was renamed as Openethereum in 2019. Openethereum is implemented using Rust programming language and is claimed to be the fastest, lightest, and most secure Ethereum client. The experiments are done on Openethereum version v3.0.0 with Rust version 1.22.1.
\end{itemize}

\begin{table*}[t!]
\caption{An overview of the benchmark used in our measurement study}\label{benchmarks}
\begin{tabular}{|c|c|c|c|c|}
\hline
 & Contract Name & Description & Language & \begin{tabular}[c]{@{}c@{}}Has 64-bit \\ Counterpart \end{tabular} \\ \hline
\multirow{6}{*}{Simple Operation} & add & Do \texttt{ADD} operation \textit{x} times & \multirow{6}{*}{Solidity \& Rust} & \multirow{6}{*}{Y} \\ \cline{2-3} 
& sub & Do \texttt{SUB} operation \textit{x} times &  &  \\ \cline{2-3} 
& shr & Do \texttt{SHR} operation \textit{x} times &  &  \\ \cline{2-3} 
& div & Do \texttt{DIV} operation \textit{x} times &  &  \\ \cline{2-3} 
& mod & Do \texttt{MOD} operation \textit{x} times &  &  \\ \cline{2-3} 
& power & Do \texttt{MUL} operation \textit{x} times &  &  \\ \hline
\multirow{3}{*}{Arithmetic} & fib & Calculate the \textit{x}th term of the Fibonacci sequence & \multirow{3}{*}{Solidity \& Rust} & \multirow{3}{*}{Y} \\ \cline{2-3} 
& matrix & Do matrix multiplication \textit{x} times &  &  \\ \cline{2-3} 
& warshall & \begin{tabular}[c]{@{}c@{}}Find shortest paths in a graph of \textit{x} nodes \\ using Floyd-Warshall algorithm \end{tabular} &  & \\ \hline
Block Status &  builtin & Read block status & Solidity & N \\ \hline
\multirow{3}{*}{Hashing} & keccak256 &  The Ethereum hashing function, keccak256 & \multirow{3}{*}{Solidity \& Rust} & \multirow{3}{*}{N} \\ \cline{2-3} 
& blake2b & \begin{tabular}[c]{@{}c@{}}Cryptographic hash function blake2b, producing \\ a 256-bit hash value, optimized with inline assembly \end{tabular} &  &  \\ \cline{2-3} 
& sha1 & \begin{tabular}[c]{@{}c@{}}Cryptographic hash function SHA-1, producing \\ a 160-bit hash value, optimized with inline assembly \end{tabular} &  &  \\ \hline
\end{tabular}
\end{table*}

\subsubsection{Selected WASM VMs}
\label{subsubsec:wasmvms}
The selected WASM VMs are introduced in the following.
\begin{itemize}
    \item[1)] \textbf{Geth WASM - Hera}~\cite{hera}.
    Geth requires EVMC support to execute WASM bytecode. The EVMC is the low-level ABI between VMs and Ethereum clients. On the VM side it supports classic EVM and ewasm. On the client-side it defines the interface for EVM implementations to access Ethereum environment and state. 
    Hera is the official Ewasm virtual machine connector. Hera is compiled as a shared library and can be dynamically loaded by Geth as its EVMC engine. Hera now supports three WASM backends: wabt (by default)\cite{wabt}, Binaryen\cite{binaryen} and WAVM\cite{wavm}. 
    The experiments are done using Hera v0.2.6 on a Geth client with EVMC v6.3.0.

    \item[2)] \textbf{Openethereum WASM - wasmi}~\cite{wasmi}.
    Openethereum currently supports two VM types - EVM and WASM. An execution engine named wasmi is integrated as a component in Openethereum. wasmi is a WASM interpreter implemented in Rust. It does not support work-in-progress WASM proposals and features that is not directly supported by the spec to guarantee correctness. In the experiment, we use wasmi v0.6.2.

    \item[3)] \textbf{EOS VM}~\cite{eosvm}. EOS VM is designed to meet the needs of EOSIO blockchains. The execution of EOS VM is deterministic and use a software implementation of float point arithmetic to ensure determinism. It implements a watchdog timer system to bound the execution time because blockchain system is resource limited. EOS VM has explicit checks and validations to ensure the security during execution. In the experiment, we use EOS v2.0.5.

    \item[4)] \textbf{Life}~\cite{life}. Life is a secure and fast WebAssembly VM built for decentralized applications, written in Go. 
    Life is an interpreter and does not rely on any native dependencies. But the correctness of Life needs to be improved.
    It passes only 66/72 of the WebAssembly execution semantics test suite. In the experiment, we use life on commit 05c0e0f.

    \item[5)] \textbf{SSVM}~\cite{ssvm}. SSVM is a high performance and enterprise-ready WASM VM for cloud, AI, and Blockchain applications. SSVM provides various tools for enabling different runtime environments such as general wasm runtime and ONNC runtime for AI model. We use SSVM v0.6.0 (interpreter mode of general wasm runtime) in the experiment.

    \item[6)] \textbf{wagon}~\cite{wagon}. wagon is a WebAssembly-based Go interpreter. It provides executables and libraries that could be embedded in Jupyter or any Go program. In the experiment, we use wagon v0.4.0.

    \item[7)] \textbf{wabt}~\cite{wabt}. wabt is a suite of tools for WebAssembly including translation between WASM text format and WASM binary format, decompilation, objdump, stack-based interpreter etc. It now supports 11 WASM proposals. In the experiment, we use wabt v1.0.16.

    \item[8)] \textbf{wasm3}~\cite{wasm3}. wasm3 is a high performance WebAssembly interpreter written in C. It has the fastest execution speed and the minimum useful system requirements. wasm3 runs on a wide range of architectures and platforms including PC, browsers, mobile phones, routers and self-hosting. In the experiment, we use wasm3 v0.4.7.

    \item[9)] \textbf{WAMR}~\cite{wamr}. WebAssembly Micro Runtime (WAMR) includes a VM core, an application framework and dynamic management of the WASM applications. WAMR supports WebAssembly interpreter, ahead of time compilation (AoT) and Just-in-Time compilation (JIT). We use the interpreter mode (WAMR-interp) and JIT mode (WAMR-jit) in the experiment. In the experiment, we use WAMR-04-15-2020.

    \item[10)] \textbf{wasmtime}~\cite{wasmtime}. wasmtime is a lightweight standalone runtime for WebAssembly. It is built on the optimizing Cranelift code generator to quickly generate high-quality machine code at runtime. In the experiment, we use wasmtime v0.15.0.
\end{itemize}

\begin{table*}[t]
\centering
\small
\caption{Information of the WebAssembly Support on Blockchain Clients}\label{table:support}
\begin{tabular}{lccc}
\hline
 & \textbf{Go-Ethereum} & \textbf{OpenEthereum (Parity)} & \textbf{EOSIO}\\ 
\hline
\hline
\multicolumn{4}{c}{\textbf{WASM Support}} \\ 
\hline
\hline
\textbf{Native Support} & N & Version 1.9.5 and later & Y\\ 
\hline
\textbf{How to Activate WASM} & \begin{tabular}[c]{@{}c@{}}Compile Geth with EVMC support \\ and launch Geth with Hera set as \\ its EVMC engine \end{tabular} & \begin{tabular}[c]{@{}c@{}}Configure \textit{wasmActivationTransition} \\ parameter in chain specification file \end{tabular} & NA \\ 
\hline
\textbf{WASM Engine} & Hera & wasmi & EOS-vm \\ 
\hline
\textbf{Status} & Unstable but usable & Unstable but usable & Stable for production usage \\ 
\hline
\hline
\multicolumn{4}{c}{\textbf{Coding Standard}} \\ 
\hline
\hline
\textbf{Contract Entry Point} & \textit{main()} & \textit{call()} & \textit{action\_wrapper} in <eosio.hpp> \\ 
\hline
\textbf{Exports} & \textit{main()} and \textit{memory} & \textit{call()} and \textit{deploy()} (optional) & \\
\hline
\textbf{Library} & NA & pwasm-ethereum & EOSIO.CDT \\ 
\hline
\hline
\multicolumn{4}{c}{\textbf{Developer Support}} \\ 
\hline
\hline
\textbf{Official Toolchain} & N & Rust nightly-2018-11-12 & EOSIO.CDT \\ 
\hline
\textbf{Binary Packer} & wasm-chisel & wasm-build & NA \\
\hline
\begin{tabular}[c]{@{}l@{}}\textbf{Language Supported} \\ \textbf{by Official Toolchain} \end{tabular} & NA & Rust & C/C++ \\ 
\hline
\textbf{Official Testnet} & Ewasm testnet & Kovan & testnet.eos.io\\
\hline
\end{tabular}
\end{table*}

\subsection{Benchmarks}

Table~\ref{benchmarks} lists the benchmark contracts used in our experiment, including 13 different kinds of benchmarks. They can be grouped into four categories according to their functionalities: \textit{simple operation}, \textit{arithmetic}, \textit{block status} and \textit{hashing}.
To better understand the performance difference of EVM bytecode and WASM bytecode, we adopt an opcode-level measurement by designing group \textit{simple operation}, in which the benchmarks differ only in key opcodes. The Solidity version of benchmarks in \textit{arithmetic}, \textit{block status} and \textit{hashing} are from the most widely used blockchain test sets: Ethereum Consensus Tests~\cite{ethereum_tests} and Blockbench~\cite{dinh2017blockbench}. We exclude the benchmarks that invoke external calls to other contracts and benchmarks that cannot be compiled correctly to WASM by SOLL compiler. We also implemented the Rust version of the corresponding benchmarks.

Benchmarks in group \textit{simple operation} execute a simple operation \textit{x} times, where \textit{x} is input value, originally set to be 10000. Benchmarks in group \textit{arithmetic} do numeric computations more complicated than \textit{simple operation}, such as Fibonacci sequence calculation and matrix multiplication. The input value \textit{x} is set to be 10000 for benchmark \textit{fib}, and 100 for benchmark \textit{matrix} and \textit{warshall} due to the stack limit of eWASM execution engines. The benchmark in group \textit{block status} read the status of current block. This benchmark cannot be executed in standalone WASM VMs, so it doesn't have Rust implementation. Benchmarks in group \textit{hashing} compute hash values using different hashing algorithms. The \textit{keccak256} invokes EVM instruction SHA3 to calculate hash values. Benchmark \textit{black2b} and \textit{sha1} implement the hashing algorithm manually and optimize the code with inline assembly\cite{black2b, sha1}.
Since EVM and eWASM do not support floating point, the benchmarks are implemented in 256-bit integer and 64-integer. Benchmarks in group \textit{block status} and \textit{hashing} only have 256-bit version because the calculation or the result cannot be fit into 64-bit integer.

\subsection{Experiment Setup}

The EVM and eWASM contracts are deployed on a 2-node private chain and executed locally without creating transactions, because we focus on the performance of VMs. The WASM contracts are executed on the standalone engines directly.

To ensure the robustness of experiment results, we run repeated experiments on three servers: one GPU server (Dual-core Intel Xeon CPU processor, two Nvidia Tesla M40 GPU, and 128 GB memory), one CPU server (8 core Intel Xeon CPU processor and 15 GB memory), and one cloud server (16 core Intel Xeon Platinum 8269 processor and 64 GB memory). All three servers are running Ubuntu 16.04. We do the experiment repeatedly for 1000 times and calculate the average running time. \textit{The results on the three servers show the same tendency and approximate relative value}. We report the experiment results on the GPU server in the following section.

\section{Measurement Study}
\label{sec:results}

In this section, we present the measurement results. We first explore the WASM supports on blockchain clients to answer \textbf{RQ1}. 
Then we present the experimental results of the performance measurement, including a comparison between EVM engines, a comparison between WASM engines, and a comparison between WASM engines and EVM engines, to answer \textbf{RQ2}, \textbf{RQ3} and \textbf{RQ4}, respectively.

\subsection{WASM Supports of Blockchain Clients} \label{sec:results_support}

Table~\ref{table:support} summarizes the WASM supports of blockchain clients.
We compare three widely used blockchain clients, including \textit{Go-Ethereum}, \textit{OpenEthereum}, and \textit{EOSIO}. The first two are Ethereum-based, while the last one is not but it supports WASM VM as its native execution engine. 
Here, we mainly compare differences of these clients from three perspectives: \textit{WASM Support}, \textit{Coding Convention} and \textit{Development Tool Support}.

\subsubsection{WASM Support}
\begin{itemize}
    \item \textbf{Native Support. } 
    First, \textbf{Geth} does not support WASM natively. The WASM support on Geth is now usable but unstable, and several EEIs are still working in progress. 
    Second, \textbf{OpenEthereum} client supports EVM and WASM from version 1.9.5 and later, with EVM being the default. 
    The WASM support on Openethereum is now usable but unstable as well. For example, the deployment and execution process of WASM contracts still encounter several unexplained exceptions~\cite{pwasm_issue}. 
    Third, \textbf{EOSIO} supports WASM VM as its native execution engine, and it provides stable WASM support for production usage.

    \item \textbf{How to activate WASM VMs. } 
    First, developers have to compile the \textbf{Geth} client with EVMC support, and launch Geth with Ewasm's official engine Hera attached as its EVMC engine. Hera leverages various WASM backends and serves as the connector between WASM VM and the Geth client. 
    Second, to activate WASM VM in \textbf{Openethereum}, the \textit{wasmActivationTransition} parameter in chain specification file should be set as non-zero. After activation, the Openetherum client will switch between the EVM and WASM VM during execution according to the first 4 bytes of the smart contract. 
    Third, \textbf{EOSIO} supports WASM VM as its native execution engine, so there is no need to activate WASM VM.
\end{itemize}

\subsubsection{Coding Convention}
\begin{itemize}
    \item \textbf{Entry point and exports.} 
    The WASM smart contracts to be deployed and executed in blockchains must follow specific format, for example, a given function be exported as the contract entry point. 
    First, \textbf{Geth} requires a contract to have exactly two exported symbols: the contract entry point \code{main} function with no parameters and no result value, and \code{memory}, the shared memory space for the EEI to write into. 
    Second, \textbf{Openethereum} requires a function named \code{call} exported as contract entry point. An optional exported function \code{deploy} will be executed on contract deployment to set the initial storage values of this contract. 
    Third, \textbf{EOSIO} exports an \textit{action} as the basic element of communication between smart contracts, so the entry point is defined by instancing template \code{action\_wrapper} with the the action name and action method as parameters. 

    \item \textbf{Library.} Specific library need to be included when developing WASM smart contracts for Openethereum and EOSIO. These libraries provide APIs for chain-dependent smart contract functionalities, provide templates for data types, and delegate the memory allocation. Since the above functionalities are provided by the eWASM runtime on Geth, there is no need to include specific library when developing smart contracts for Geth.
\end{itemize}

\subsubsection{Development Tool Support}
\begin{itemize}
    \item \textbf{Toolchain. } 
    First, \textbf{Geth} does not provide official WASM toolchain for developers. 
    Second, \textbf{Openethereum} provides an official tutorial that uses Rust nightly-2018-11-12 as the toolchain. 
    Third, as the official toolchain of \textbf{EOSIO}, EOSIO.CDT (i.e., Contract Development Toolkit) provides a set of tools and libraries to facilitate smart contract development for the EOSIO platform. 

    \item \textbf{Binary packer.} After compiling the source code to WASM bytecode, Geth and Openethereum requires a binary packer to transform WASM into a restricted format that can be used in blockchain context.

    \item \textbf{Testnet. } All three clients have official testnets.
\end{itemize}

\begin{framed}
\noindent \textbf{Answer to RQ1:} 
\textit{
The support for WASM on Ethereum-based blockchain clients is far from ideal. 
First, the WASM support on Ethereum-based blockchain clients is experimental and not stable. A number of configuration efforts are required to activate the WASM-verison engines for both Geth and OpenEthereum.
Second, the WASM coding convention varies across Ethereum blockchain clients, which makes it impossible to implement WASM smart contracts in a general form, i.e., developers have to write and compile smart contracts separately for different clients. Third, few languages are supported by existing tool chains, and the contract development process is cumbersome and error-prone.
}
\end{framed}

\subsection{A Comparison of EVM Engines}

To measure the performance gap between EVM engines of different blockchain clients, we deploy the 13 benchmark contracts on Geth and Openethereum. 
We gather the execution time of each client executing the benchmark contracts and calculate the average for 1,000 runs. 
After that, we calculate the \texttt{time ratio}, i.e., \texttt{T(Openethereum) / T(Geth)} for both 256-bit and 64-bit versions.
Table~\ref{table:evm} presents the overall results. Note that NA means that benchmark does not have a 64-bit version.

\begin{table}[htb]
\caption{Execution time ratio of Openethereum to Geth. 
\textit{NA} indicates that the benchmark does not have a 64-bit version.}\label{table:evm}
\begin{tabular}{|c|r|r|r|}
\hline
\textbf{Benchmark} & \begin{tabular}[c]{@{}c@{}}\textbf{Time ratio} \\ \textbf{(256-bit)}\end{tabular} & \begin{tabular}[c]{@{}c@{}}\textbf{Time ratio} \\ \textbf{(64-bit)}\end{tabular} & \textbf{64-bit speedup}\\
\hline
add & 5.545 & 5.379 & 1.03\\
sub & 5.445 & 5.248 & 1.04\\
shr & 5.111 & 4.975 & 1.03\\
div & 4.949 & 4.344 & 1.14\\
mod & 4.816 & 4.814 & 1.00\\
power & 5.038 & 4.916 & 1.02\\
\hline
fib & 5.113 & 5.204 & \textbf{0.98}\\
matrix & 3.149 & 4.932 & \textbf{0.64}\\
warshall & 4.757 & 4.629 & 1.03\\
\hline
builtin & 5.503 & NA & NA\\
\hline
keccak256 & 3.002 & NA & NA\\
blake2b & \textbf{0.927} & NA & NA\\
sha1 & \textbf{1.159} & NA & NA\\
\hline
\end{tabular}
\end{table}

\begin{figure}[t!]
	\includegraphics[width=0.99\linewidth]{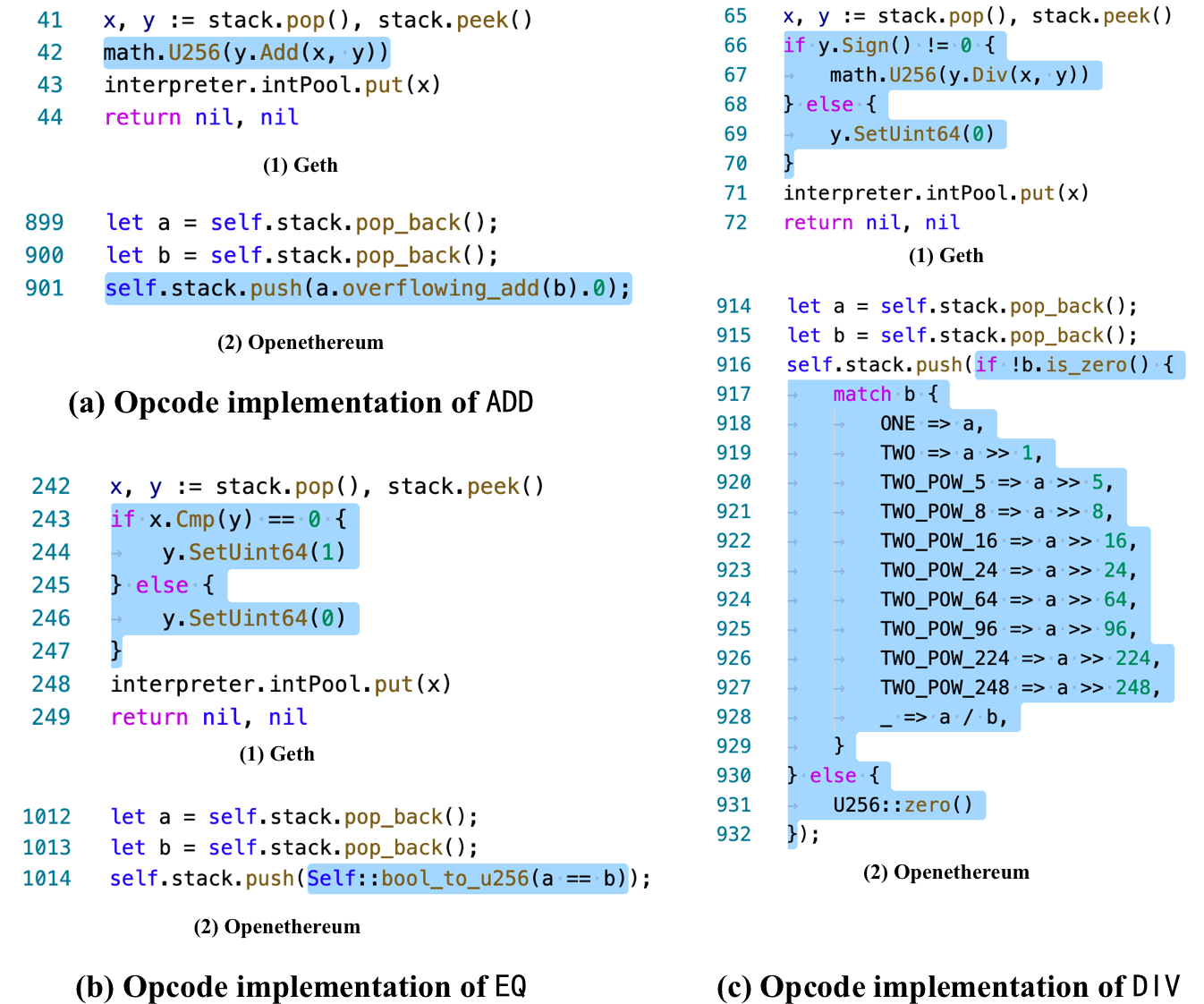}
	\caption{Examples of opcode implementation in Geth and Openethereum.}
	\label{figure:opcodes}
\end{figure}

\subsubsection{Overall Performance Comparison}
In general, Geth EVM engine runs faster than Openethereum EVM engine for most of the cases. 
The \texttt{time ratio} of Openethereum to Geth is between 3 and 6 in group \textit{simple operation}, \textit{arithmetic} and \textit{block status}, indicating that Geth EVM engine runs 3 to 6 times faster than Openethereum EVM engine, regardless of the integer type (64-bit and 256-bit) of the smart contract. 
The \texttt{time ratio} in group \textit{hashing} is smaller than other groups. When the source code is optimized manually with inline assembly, the performance gap between the two EVMs is significantly reduced. The time ratio is 0.927 and 1.159 on benchmark \textit{blake2b} and \textit{sha1} respectively.

To explore the reason why Openethereum runs slower, we manually analyzed the source code of Geth EVM and Openethereum EVM, and found that the different implementation of opcode might be the leading reason. 
As shown in Figure~\ref{figure:opcodes}, Geth and Openethereum implement the same opcodes in quite different ways. 
For example, the opcode \texttt{ADD} (Figure~\ref{figure:opcodes}(a)) is implemented with a simple \textit{Add} operation in Geth, however, Openethereum will further check if an arithmetic overflow occurs. Figure~\ref{figure:opcodes}(b) shows another case: the opcode \texttt{EQ} is implemented by an if-else condition and a 64-bit result is pushed directly into the stack. While in Openethereum, a function named \textit{bool\_to\_u256} is called to transform a boolean into 256-bit integer result. Figure~\ref{figure:opcodes}(c) shows the different implementation of opcode \texttt{DIV}. Geth implements the opcode \texttt{DIV} with simple division operation. Openethereum leverages the shift right operation to optimize the execution with a fall back to the primitive division operation. However, if the divisor does not match the given value, the calculation will go through 10 more checks, leading to a worse performance. Thus, it is apparently that the opcode-level implementation of EVM engines would greatly impact the overall performance of smart contract execution.

\subsubsection{64-bit VS. 256-bit}
The column 4 in Table~\ref{table:evm} compares the \texttt{time ratio} of 256-bit smart contract to that of 64-bit version, i.e., \texttt{T(256-bit) / T(64-bit)}.
Obviously, smart contracts with different data types have shown no performance gap when running on the EVM engines. \textit{The main reason is that the native data type of EVM is 256-bit integer, and it supports down to 8-bit integer calculation. Thus, switching the data type does not introduce any obvious impact to the execution performance of smart contracts on EVM engines.}

\begin{framed}
\noindent \textbf{Answer to RQ2:} 
\textit{
For the two most popular Ethereum EVM engines, the performance of contract execution on Geth is better than that on Openethereum. Our investigation suggests that the opcode-level implementation of EVM engines would greatly impact the overall performance of smart contract execution. Moreover, 64-bit and 265-bit smart contract have shown no obvious performance gap during their execution on EVM engines, as the native data type of EVM bytecode is 256-bit.
}
\end{framed}

\subsection{A Comparison of WASM Engines}

We compile the Rust version of 12 benchmarks (in group \textit{simple operation}, \textit{arithmetic} and \textit{hashing}) into WASM bytecode and execute the bytecode on 11 standalone WASM VMs: \texttt{WAVM} (one backend of Geth-Hera), \texttt{wasmi}, \texttt{EOS VM}, \texttt{Life}, \texttt{SSVM}, \texttt{wagon}, \texttt{wabt}, \texttt{wasm3}, \texttt{WAMR-interp}, \texttt{WAMR-jit} and \texttt{wasmtime}. 
Note that, we use two different modes of \texttt{WAMR} (\texttt{WAMR-interp} and \texttt{WAMR-jit}) for evaluation.
We gather the execution time of executing the benchmark for each standalone WASM engine, and calculate the average of 1,000 runs.

\subsubsection{Overall Performance Comparison.}
Table~\ref{table:wasm} shows the execution time of our benchmark across the 11 types of standalone WASM engines. 
Surprisingly, the performance across standalone WASM engines differs greatly. \textit{For different benchmark, there is up to three orders of magnitude difference between the performance of the best engine and the worst engine.} For example, it takes over 19,768 ms for the \texttt{EOS VM} to execute the \textit{matrix} benchmark, while it only takes roughly 70ms for \texttt{WAVM} and \texttt{wasmtime} engines.
In general, wasm3 runs the fastest of all WASM VMs, with execution time less than 1ms for most of the cases. 
However, \texttt{WAVM}, the engine used in the WASM-version client of Geth, achieves the worst result for all of the 64-bit benchmarks. 
For some specific 256-bit versions of benchmarks including \textit{shr}, \textit{div}, \textit{mod}, \textit{fib} and \textit{matrix}, EOS VM takes the longest execution time. The execution time of the remaining VMs is between the execution times of the above three engines. We speculate that the performance gap is introduced by the diverse implementations of engines.
\textit{This result suggests that, although there are a number of WASM engines available, not all of them can be used to achieve excellent performance, especially when adopting WASM to the blockchain environment.}

\subsubsection{64-bit VS. 256-bit}
From a vertical analysis (the two rows for each benchmark contract), we can measure the impact of integer types on the performance of WASM bytecode execution. \textit{It is interesting to see that, the integer types in the contract will significantly affect the performance of WASM bytecode execution, across all the engines we studied.}
In general, 256-bit benchmarks run slower than their 64-bit counterparts. Most of the WASM engines gain more than 2x speedup for their 64-bit versions.
For the most representative benchmarks including \textit{matrix}, \textit{fib} and \textit{mod}, the execution of 64-bit smart contract can achieve great speedup for all the engines. For example, it takes roughly 1514ms to execute the 256-bit version of \textit{fib} contact on the EOS VM, while the time can be reduced to 1.69ms for its 64-bit version.
It can be explained that, WASM bytecode and EVM bytecode support different types of integers. As the comparison in Table~\ref{table:ewasm_design}, WASM uses 32/64-bit bytecode but EVM supports 256-bit bytecode. 
For the 256-bit smart contract, it must be converted before the WASM engine can execute them, which would introduce additional overhead.

\begin{table*}[t!]
\caption{Execution time (ms) of standalone WASM VMs}\label{table:wasm}
\small
\begin{tabular}{|c|c|r|r|r|r|r|r|r|r|r|r|r|}
\hline
\textbf{Benchmark} &  & \textbf{WAVM} & \textbf{wasmi} & \textbf{EOS VM} & \textbf{Life} & \textbf{SSVM} & \textbf{wagon} & \textbf{wabt} & \textbf{wasm3} & \begin{tabular}[c]{@{}c@{}}\textbf{WAMR}\\\textbf{-interp}\end{tabular} & \begin{tabular}[c]{@{}c@{}}\textbf{WAMR}\\\textbf{-jit}\end{tabular} & \textbf{wasmtime}\\
\hline
\multirow{2}{*}{add} & 256-bit & 11.51 & 2.85 & 13.19 & 4.76 & 2.21 & 2.88 & 2.32 & 0.13 & 1.40 & 1.37 & 0.59\\
\cline{2-13} 
& 64-bit & 10.12 & 1.07 & 1.78 & 2.50 & 1.09 & 1.25 & 1.93 & 0.04 & 1.20 & 1.20 & 0.42\\
\hline
\multirow{2}{*}{sub} & 256-bit & 11.54 & 2.80 & 13.86 & 4.62 & 2.21 & 2.85 & 2.35 & 0.13 & 1.38 & 1.40 & 0.58\\
\cline{2-13} 
& 64-bit & 10.07 & 1.08 & 1.75 & 2.56 & 1.10 & 1.26 & 1.93 & 0.04 & 1.18 & 1.20 & 0.42\\
\hline
\multirow{2}{*}{shr} & 256-bit & 14.76 & 31.21 & 161.78 & 28.74 & 16.04 & 24.67 & 9.38 & 1.28 & 5.44 & 5.57 & 4.44 \\
\cline{2-13} 
& 64-bit & 10.12 & 1.08 & 1.60 & 2.66 & 1.05 & 1.29 & 1.96 & 0.04 & 1.19 & 1.21 & 0.45 \\
\hline
\multirow{2}{*}{div} & 256-bit & 76.67 & 73.30 & 355.31 & 73.65 & 40.23 & 74.25 & 17.27 & 4.27 & 14.15 & 14.14 & 27.03 \\
\cline{2-13} 
& 64-bit & 10.10 & 1.08 & 1.54 & 2.64 & 1.09 & 1.29 & 1.95 & 0.04 & 1.16 & 1.20 & 0.44\\
\hline
\multirow{2}{*}{mod} & 256-bit & 82.83 & 140.27 & 902.68 & 161.73 & 81.29 & 156.97 & 31.35 & 10.02 & 24.60 & 24.63 & 28.11\\
\cline{2-13} 
& 64-bit & 10.10 & 1.07 & 1.67 & 2.70 & 1.08 & 1.28 & 1.89 & 0.04 & 1.19 & 1.18 & 0.47\\
\hline
\multirow{2}{*}{power} & 256-bit & 10.12 & 1.09 & 1.83 & 2.70 & 1.09 & 1.26 & 1.96 & 0.04 & 1.21 & 1.22 & 0.46\\
\cline{2-13} 
& 64-bit & 10.13 & 1.06 & 1.67 & 2.48 & 1.07 & 1.22 & 1.94 & 0.04 & 1.18 & 1.19 & 0.41\\
\hline
\multirow{2}{*}{fib} & 256-bit & 86.68 & 241.16 & 1514.43 & 266.32 & 146.09 & 260.24 & 55.45 & 13.61 & 45.22 & 45.24 & 31.40\\
\cline{2-13} 
& 64-bit & 10.12 & 1.09 & 1.69 & 2.64 & 1.07 & 1.29 & 1.90 & 0.04 & 1.21 & 1.15 & 0.46\\
\hline
\multirow{2}{*}{matrix} & 256-bit & 70.43 & 2354.51 & 19768.51 & 3225.61 & 1136.63 & 2813.39 & 407.37 & 132.42 & 304.03 & 303.63 & 72.08\\
\cline{2-13} 
& 64-bit & 18.06 & 23.32 & 116.54 & 22.31 & 12.86 & 20.75 & 7.92 & 1.23 & 5.24 & 5.18 & 4.63\\ 
\hline
\multirow{2}{*}{warshall} & 256-bit & 90.50 & 12.30 & 40.65 & 19.94 & 6.33 & 33.74 & 4.41 & 0.44 & 2.43 & 2.40 & 48.06\\
\cline{2-13} 
& 64-bit & 28.25 & 3.74 & 9.44 & 5.70 & 1.81 & 11.95 & 2.54 & 0.10 & 1.30 & 1.33 & 9.67 \\
\hline
keccak256 & - & 12.84 & 2.13 & 2.42 & 2.89 & 1.11 & 1.78 & 1.93 & 0.05 & 1.19 & 1.18 & 1.13\\
\hline
blake2b & - & 112.23 & 3.70 & 8.12 & 17.13 & 1.76 & 35.19 & 2.85 & 0.42 & 1.23 & 1.18 & 21.91\\
\hline
sha1 & - & 83.05 & 3.15 & 5.88 & 13.11 & 1.50 & 26.79 & 2.58 & 0.33 & 1.22 & 1.20 & 17.59\\
\hline
\end{tabular}
\end{table*}

\begin{framed}
\noindent \textbf{Answer to RQ3:} 
\textit{
The performance of smart contract execution across standalone WASM engines differs greatly, even up to three orders of magnitude difference for some contracts. It suggests that not all the popular WASM engines can be used to achieve excellent performance. Furthermore, the native data type has a great impact on the performance of smart contract execution, as the native data type of WASM is 32/64 bit, it would introduce additional overhead during the execution of 256-bit smart contracts.
}
\end{framed}

\subsection{WASM Engines VS. EVM Engines}

As aforementioned, we compile the Solidity smart contracts into both EVM bytecode and eWASM bytecode. Then we deploy the contracts on Geth and Openethereum, and gather the execution time of EVM and WASM engines on the same blockchain client.

\begin{table*}[t!]
\caption{A Comparison of eWASM and EVM bytecode execution on Geth and Openethereum.}\label{table:evm_wasm}
\small
\begin{tabular}{|c|c|r|r|r|r|r|r|}
\hline
\multicolumn{2}{|c|}{\multirow{2}{*}{\textbf{Benchmark}}} & \multicolumn{3}{c|}{\textbf{Geth}} & \multicolumn{3}{c|}{\textbf{Openethereum}}\\
\cline{3-8} 
 \multicolumn{2}{|c|}{} & \begin{tabular}[c]{@{}c@{}}\textbf{Overall} \\ \textbf{Performance}\end{tabular} & \begin{tabular}[c]{@{}c@{}}\textbf{Execution} \\ \textbf{Overhead}\end{tabular} & \begin{tabular}[c]{@{}c@{}}\textbf{Net} \\ \textbf{Performance}\end{tabular} & \begin{tabular}[c]{@{}c@{}}\textbf{Overall} \\ \textbf{Performance}\end{tabular} & \begin{tabular}[c]{@{}c@{}}\textbf{Execution} \\ \textbf{Overhead}\end{tabular} & \begin{tabular}[c]{@{}c@{}}\textbf{Net} \\ \textbf{Performance}\end{tabular} \\
\hline
\multirow{2}{*}{add} & 256-bit & 25.07 & 92.8\% & 1.81 & 4.76 & 96.0\% & 0.19\\
\cline{2-8} 
& 64-bit & 16.86 & 93.0\% & 1.18 & 0.17 & 87.8\% & 0.02\\
\hline
\multirow{2}{*}{sub} & 256-bit & 23.25 & 93.5\% & 1.51 & 4.59 & 96.0\% & 0.18\\
\cline{2-8} 
& 64-bit & 21.21 & 93.9\% & 1.29 & 0.28 & 87.7\% & 0.03\\
\hline
\multirow{2}{*}{div} & 256-bit & 32.52 & 78.3\% & 7.06 & 64.21 & 96.3\% & 2.38\\
\cline{2-8} 
& 64-bit & 9.99 & 93.1\% & 0.69 & 3.35 & 98.4\% & 0.05\\
\hline
\multirow{2}{*}{mod} & 256-bit & 109.33 & 92.3\% & 8.41 & 111.85 & 96.7\% & 3.69\\
\cline{2-8} 
& 64-bit & 14.45 & 93.8\% & 0.90 & 2.95 & 98.3\% & 0.05\\
\hline
\multirow{2}{*}{power} & 256-bit & 101.09 & 98.8\% & 1.21 & 59.65 & 99.9\% & 0.06\\
\cline{2-8} 
& 64-bit & 16.73 & 92.8\% & 1.20 & 0.12 & 88.0\% & 0.01\\
\hline
\multirow{2}{*}{fib} & 256-bit & 138.33 & 95.8\% & 5.81 & 171.02 & 97.5\% & 4.28\\
\cline{2-8} 
& 64-bit & 9.09 & 93.1\% & 0.63 & 2.10 & 98.5\% & 0.03\\
\hline
\multirow{2}{*}{matrix} & 256-bit & 3.95 & 33.6\% & 2.62 & 85.19 & 97.4\% & 2.21\\
\cline{2-8} 
& 64-bit & 4.06 & 30.7\% & 2.81 & 13.96 & 99.8\% & 0.03\\
\hline
\multirow{2}{*}{warshall} & 256-bit & 5.65 & 61.7\% & 2.16 & 20.92 & 94.6\% & 1.13\\
\cline{2-8} 
& 64-bit & 4.80 & 58.8\% & 1.98 & 9.36 & 95.8\% & 0.39\\
\hline
keccak256 & 256-bit & 33.84 & 98.5\% & 0.51 & 119.80 & 87.1\% & 15.45\\
\hline
\end{tabular}
\end{table*}

\subsubsection{Overall Performance Comparison}
Table~\ref{table:evm_wasm} shows the execution time ratio of WASM to EVM (\texttt{T(WASM) / T(EVM)}.) in executing benchmark contracts with different integer types. A larger value indicates that the WASM engine is slower than the EVM engine on the same blockchain client. The statistics show a counter-intuitive result. The eWASM engines are slower than EVM engines for all of the 256-bit benchmarks and most of the 64-bit benchmarks on both Geth and Openethereum clients. 
When the integer type is 256-bit, on Geth client (see column 3 on Table~\ref{table:evm_wasm}), the time ratio of WASM VM to EVM is between 3.95 (benchmark \textit{matrix}) and 138.33 (benchmark \textit{fib}). The time ratio of benchmark \textit{mod}, \textit{power}, and \textit{fib} is significantly larger than others. 
It may because all of these smart contracts involve \texttt{MUL} and \texttt{MOD} instructions. Benchmark \textit{matrix} and \textit{warshall} have significantly smaller time ratio than other benchmarks, This may because these two contracts involve multiple \texttt{MLOAD} and \texttt{MSTORE} instructions related to memory operating. On Openethereum client, the time ratio of WASM VM to EVM is between 4.59 (benchmark \textit{sub}) and 171.02 (benchmark \textit{fib}).

\subsubsection{64-bit VS. 256-bit}
The time ratio of WASM VM to EVM of 64-bit benchmarks is significantly lower than their 256-bit counterpart. As shown in Table~\ref{table:evm_wasm_speedup}, switching from 256-bit to 64-bit achieves a speedup up to 15x and 497x on Geth and Openethereum, respectively. When executing benchmark \textit{add}, \textit{sub} and \textit{power} on Openethereum, WASM VM even runs faster than EVM. This is because the native data type of WASM is 32/64-bit, thus computations higher than 64-bit must be converted to a 256-bit format before the WASM engine can process them. While 64-bit contracts do not go through type conversion, which leads to a faster execution.

\begin{table}[h]
\caption{Overall performance speedup of 64-bit benchmarks to corresponding 256-bit benchmarks on Geth and Openethereum.}
\label{table:evm_wasm_speedup}
\begin{tabular}{|c|r|r|}
\hline
\textbf{Benchmark} & \textbf{Geth} & \textbf{Openethereum} \\
\hline
add & 1.49x & 28x \\
\hline
sub & 1.10x & 16.39x\\
\hline
div & 3.26x & 19.17x\\
\hline
mod & 7.57x & 37.92x\\
\hline
power & 6.04x & 497.08x\\
\hline
fib & 15.22x & 81.44x\\
\hline
matrix & 0.97x & 6.10x\\
\hline
warshall & 1.18x & 2.24x\\
\hline
\end{tabular}
\end{table}

\subsubsection{Understanding the Execution Overhead}
As aforementioned in Section~\ref{sec:background_ewasm}, to be executed in the Ethereum environment, eWASM interacts with the blochain through EEI methods, restricts non-deterministic behavior, and conducts gas metering before the execution of WASM bytecode. Compared to the WASM execution on native engines, the above operations introduce additional overhead.

\begin{figure*}[htb]
    \centering
    \subfloat[Geth]
    {
        \begin{minipage}[t]{0.48\linewidth}
            \centering
            \includegraphics[width=0.95\linewidth]{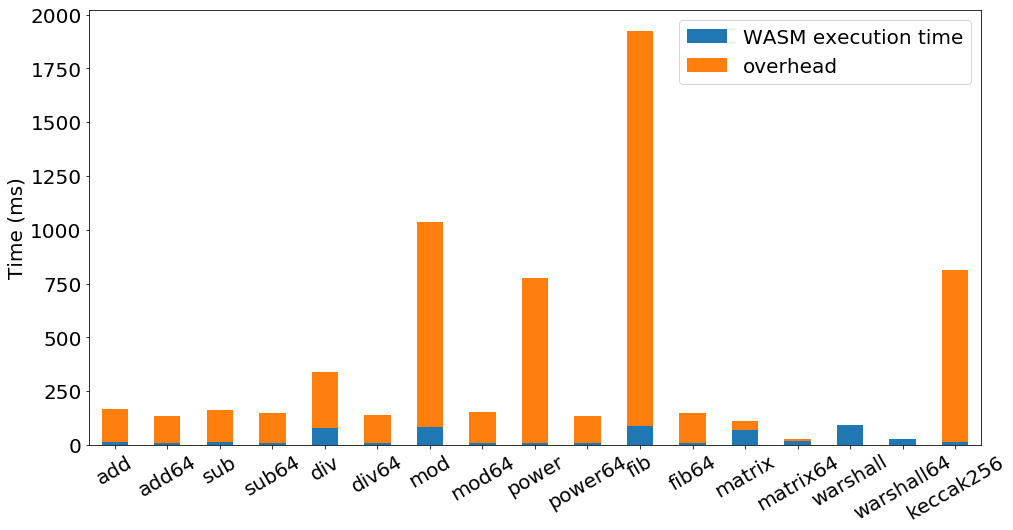}
            \label{geth_evm_wasm_overhead}
        \end{minipage}
    }
    \subfloat[Openethereum]
    {
        \begin{minipage}[t]{0.48\linewidth}
            \centering
            \includegraphics[width=0.95\linewidth]{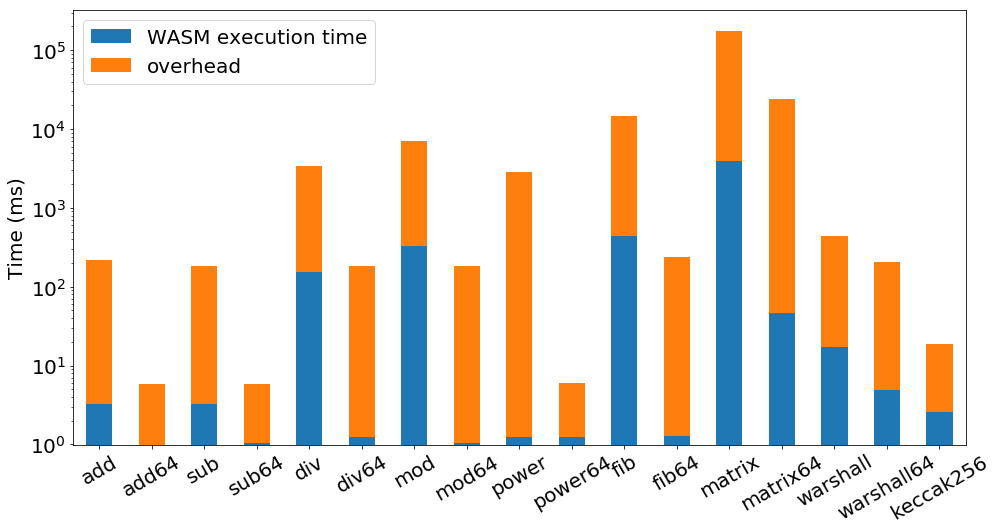}
            \label{parity_evm_wasm_overhead}
        \end{minipage}
    }
    \caption{Execution overhead of WASM VM running on Blockchain clients}
    \label{evm_wasm_overhead}
\end{figure*}

\textbf{Measuring the Overhead.}
To measure the execution overhead of WASM VMs in the blockchain environment, we compare the performance of smart contract execution on eWASM engine with the results on the corresponding standalone WASM engines.
We use the difference of the result \texttt{T(overhead) = T(eWASM) - T(standalone WASM)} to approximate the execution overhead of eWASM on Ethereum clients. Table~\ref{table:evm_wasm} shows the statistics of the results. As shown in Figure~\ref{geth_evm_wasm_overhead}, on Geth, the execution overhead is far more than the execution time of standalone WASM VM. The execution overhead takes up 84.9\% of WASM execution time on average, with the maximum overhead taking up 98.8\% of total execution time (benchmark \textit{power}). Figure~\ref{parity_evm_wasm_overhead} is shown in log scaled. Openethereum exhibits a similar result to Geth. The execution overhead takes up 95.0\% of WASM execution time on average, with the maximum overhead taking up 99.9\% of total execution time (benchmark \textit{power}).

\textbf{Excluding the Overhead.}
Table~\ref{table:evm_wasm} shows the execution time ratio of WASM to EVM after excluding the overhead, which can be expressed as:
\texttt{T(standalone WASM) / T(EVM)}.
Interestingly, without considering the overhead, for most cases, the performance of eWASM engine is superior to that of WASM engine on Openethereum, and the performance eWASM engine is quite close to that of WASM engine on Geth.
According to the design of eWASM, the execution overhead comes from gas metering before the execution of WASM bytecode, the deterministic stack height metering and the EEI methods. This also explains the performance gap between 256-bit contracts and 64-bit contracts. Since the conversion of integer types are done during compilation, the compiled WASM bytecode for 256-bit contracts contain more instructions, so more overhead is introduced.

\begin{framed}
\noindent \textbf{Answer to RQ4:} 
\textit{While standalone WASM VMs are faster than EVMs, the blockchain flavored WASM VMs are slower than EVMs for all of the 256-bit benchmarks and most of the 64-bit benchmarks on both Geth and Openethereum clients. This is because of the overhead introduced by gas metering before the execution of WASM bytecode and the inefficient context switch of EEI methods. }
\end{framed}

\section{Discussion}\label{sec:discussion}
\subsection{Implication}
Our findings reveal that the support for WASM on Ethereum-based blockchain clients is far from ideal, including the experimental and unstable WASM support, non-generic coding conventions and insufficient support of existing tool chains, which cause obstacles to the spreading of WASM.

Our observations are also helpful to stakeholders in the blockchain community, including blockchain client developers, VM developers and compiler developers. For blockchain client, our results can help developers optimize them by reducing additional overhead, including the execution process of WASM on the blockchain client and the interface design between WASM and the blockchain environment. For the virtual machine, our results can help developers improve the performance of VM by removing/adjusting bottleneck instructions. For the compiler, Our result can help developers refine the implementation of compiler by considering the performance gap between 256-bit version and 64-bit one. 
Furthermore, We implement a relatively comprehensive benchmark set with various integer types and different input in Solidity and Rust. 
We will release the benchmark set to the community, and we believe that it can benefit future research on performance measurement for EVM and WASM VM.

\subsection{Limitation}

Our study, however, carries several limitations. 
First, our measurement study is limited by the benchmark we created. According to the design of smart contract~\cite{szabo1997formalizing,ethereum}, the smart contracts execute specific code logic, manage their own ether balance and key/value store, and can transact with other contracts. In our measurement study, we have implemented 13 different kinds of benchmarks, that perform simple operation, arithmetic computation, block status checking and hashing. We do not include benchmarks that require long-term storage such as key/value store operating, because the standalone WASM engines do not support long-term storage. Moreover, We do not include benchmarks that involve transactions or interaction with other contracts, as the performance of transactions can be affected by many factors, which is beyond the scope of this research. Related work considering blockchain transaction performance can be found in ~\cite{dinh2017blockbench, zheng2018detailed, rouhani2017performance}.
Second, we design a group of benchmarks to approximately measure the execution of opcodes. A more precise approach is instrumenting the execution engine to record the execution time of opcodes. However, instrumenting the opcode execution will introduce a relatively large overhead. We compare the blockchain flavoured WASM results with their corresponding standalone WASM VM results and use the difference of the result to approximate the execution overhead of Ethereum flavoured WASM. We will measure the detailed volume and location of eWASM overhead in future work.
Third, the WASM support of blockchain clients is in an experimental stage which evolves quickly. The WASM VMs are releasing updates frequently. Thus, the results of this study may not reflect the up-to-date performance of the VMs. To reduce the bias, we ensure the blochchain clients and standalone WASM VMs were the latest release before July 2020. At last, in this study, we only consider the execution performance of smart contract bytecode itself, while the actual performance of smart contract invocation in the blockchain environment relies on other factors such as consensus protocols. Nevertheless, our measurement study in this paper is performed on the clients that support different types of bytecode, which can accurately reflect the net improvement introduced by bytecode VMs.

\section{Related Work}\label{sec:related}
To the best of our knowledge, this paper is the first study to characterize the impact of WASM and EVM engines on the execution performance of smart contracts. Nevertheless, there are some related work on the general in-broswer WASM study and the performance measurement of Blockcahin systems.

\subsection{WASM Bytecode Analysis}\tab
WASM is increasingly popular in recent years. Musch et al.~\cite{musch2019new} studied the prevalence of WebAssembly in the Alexa Top 1 million websites and found that 1 out of 600 sites use WebAssembly by then. The applications of WebAssembly spread over mining, obfuscation games, custom code, libraries, and environment fingerprinting techniques. 
Rüth et al.~\cite{ruth2018digging} found that the majority of in-browser miners utilizes WebAssembly for efficient PoW calculation. Although the use of WebAssembly might lead to the problem of in-browser cryptojacking, several detection methods have been proposed. Bian et al.~\cite{bian2020minethrottle} proposed a browser-based defense mechanism against WebAssembly cryptojacking. 
Lehmann et al.~\cite{lehmann2019wasabi} proposed a dynamic analysis tool named Wasabi to meet the demand of analyzing WebAssembly.
An ecosystem of WebAssembly is gradually growing.

\subsection{Performance Measurement of WASM}
Haas et al.~\cite{haas2017bringing} measured the execution time of the PolyBenchC benchmarks running on WebAssembly on V8 and SpiderMonkey normalized to native execution. The results show that WebAssembly is very competitive with native code, with nearly all of the benchmarks within 2× of native. They also showed that the on average, WebAssembly code is 62.5\% the size of asm.js, and 85.3 \% of native x86-64 code size.
Herrera et al.~\cite{herrera2018webassembly} further evaluated the performance of JavaScript and WebAssembly using the Ostrich benchmark suite, and demonstrated significant performance improvements for WebAssembly, in the range of 2x speedups over the same browser’s JavaScript engine. They also evaluated the relative performance of portable versus vendor-specific browsers, and the relative performance of server-side versus client-side WebAssembly.
Jangda et al.~\cite{jangda2019not} expanded the benchmark test suite to SPEC CPU suite. They built BROWSIX-WASM to run unmodified WebAssembly-compiled Unix applications directly inside the browser and found that applications compiled to WebAssembly run slower by an average of 45\% (Firefox) to 55\% (Chrome) compared to native. The root causes of this performance gap is that the instructions produced by WebAssembly have more loads/stores and more branches.

\subsection{Performance Measurement of Blockchain}
Previous work focused on measuring the performance of the overall Blockchain transitions.
Dinh et al.~\cite{dinh2017blockbench} proposed an evaluation framework for analyzing private blockchains named BLOCKBENCH and conducted an evaluation of Ethereum, Parity and Hyperledger. They revealed that the main bottlenecks are the consensus protocols in Hyperledger and Ethereum, but transaction signing for Parity. They also revealed that the execution engine of Ethereum and Parity is less efficient than that of Hyperledger.
Zheng et al.~\cite{zheng2018detailed} proposed a performance monitoring framework, overall performance metrics and detailed performance metrics for the users to know the exact performance in different stages of the blockchain. They observed that the blockchain with PoW consensus protocol shows lower overall performance than others. They further analyzed the performance of different smart contracts.
Rouhani et al.\cite{rouhani2017performance} analyzed the detailed performance of transitions on two most popular Ethereum clients, Geth and Parity, and showed that the transactions are 89.8\% on average faster in Parity client in comparison with Geth client under the same system configuration.
Lee et al.~\cite{lee2020measurements} studied the graph properties over Ethereum blockchain networks and subnetworks by analyzing the interactions the traces of transitions and calls of smart contracts. They found that these blockchain networks are very different from social networks but are similar to the Web. Blockchain networks are small-world and well-connected.

\subsection{Moving WASM to Blockchains.}

Ethereum WebAssembly (\emph{eWASM})~\cite{ewasmdesign} is a proposed redesign of the Ethereum smart contract execution layer using a deterministic subset of WebAssembly. \emph{eWASM} specifies the VM semantics, contract semantics, an Ethereum environment interface to interaction with, system contracts and metering for instructions.
SOLL~\cite{soll} is a new compiler for generating eWASM files from Solidity and Yul based on LLVM. The current version of SOLL can pass 34\% of Solidity compilation tests and 79\% of Yul compilation tests.

\section{Conclusion}\label{sec:conclusion}
In this work, we have presented the first measurement study to date, on the performance of smart contract execution across EVM and WASM VM. To pinpoint the root cause leading to the performance bottleneck, we provide a comprehensive study that includes both intra-byte code engine comparison (i.e., EVM vs. EVM, WASM vs. WASM), and inter-bytecode engine comparison (i.e., EVM-version engine VS. WASM-version engine of the same blockchain client). Our exploration reveals a number of issues in adopting WASM to blockchain, and our findings can provide insightful implications to the future design, implementation, and optimization of blockchain clients, WASM engines and compilers.

\balance 

\bibliographystyle{ACM-Reference-Format}
\bibliography{dlob}

\end{document}